\documentclass{vgtc}                          

\graphicspath{{xzh_img/}{imgs/}} 

\usepackage{times}                     

\usepackage{tabu}                      
\usepackage{booktabs}                  
\usepackage{lipsum}                    
\usepackage{mwe}                       

\usepackage{mathptmx}                  

\usepackage{verbatim}

\usepackage[normalem]{ulem}

\usepackage{float}

\onlineid{1182}

\vgtccategory{Research}

\vgtcinsertpkg

\newcommand{\system}{{\rmfamily \scshape vitaLITy~2}}

\newcommand{\prevsystem}{{\rmfamily \scshape vitaLITy~1}}

\newcommand{\userA}{Aaron}
\newcommand{\userN}{Noori}

\title{\system{}: Reviewing Academic Literature\\Using Large Language Models}

\author{Hongye An\thanks{e-mail: psxah15@exmail.nottingham.ac.uk}\\ %
        \scriptsize University of Nottingham %
\and Arpit Narechania\thanks{e-mail: arpitnarechania@gatech.edu}\\ %
     \scriptsize Georgia Institute of Technology
\and Emily Wall\thanks{e-mail: emily.wall@emory.edu}\\ %
     \scriptsize Emory University
\and Kai Xu\thanks{e-mail: kai.xu@nottingham.ac.uk}\\ %
     \scriptsize University of Nottingham %
}

\abstract{
Academic literature reviews have traditionally relied on techniques such as keyword searches and accumulation of relevant back-references, using databases like Google Scholar or IEEEXplore.
However, both the precision and accuracy of these search techniques is limited by the presence or absence of specific keywords, making literature review akin to searching for needles in a haystack. 
We present \system{}, a solution that uses a Large Language Model or LLM-based approach to identify semantically relevant literature in a textual embedding space. 
We include a corpus of 66,692 papers from 1970-2023 which are searchable through text embeddings created by three language models. 
\system{} contributes a novel Retrieval Augmented Generation (RAG) architecture and can be interacted with through an LLM with augmented prompts, including summarization of a collection of papers. 
\system{} also provides a chat interface that allow users to perform complex queries without learning any new programming language. This also enables users to take advantage of the knowledge captured in the LLM from its enormous training corpus. 
Finally, we demonstrate the applicability of \system{} through two usage scenarios.
\system{} is available as open-source software at \url{https://vitality-vis.github.io}.}


\begin{document}


\firstsection{Introduction}

\maketitle
In recent years, the proliferation of vast digital repositories of academic papers has posed numerous challenges for researchers and scholars alike, including (i) efficiently retrieving pertinent information from this expansive pool, (ii) comprehensively visualizing the relationships within scholarly literature, and (iii) summarizing extensive bodies of literature.
Traditional information retrieval methods for literature reviews often fall short in capturing the nuanced connections between academic papers, thereby hindering the seamless extraction of relevant insights; e.g., existing literature review approaches are often ad hoc rather than systematic~\cite{snyder2019literature} and even systematic approaches using techniques like keyword search can lead to inadequate coverage of evidence~\cite{baumeister1997writing,snyder2019literature}.

\prevsystem{} began to address these challenges by introducing a transformer-based approach to exploring academic literature~\cite{narechania_vitality_2022}. This system included a web-based visualization interface that used text embeddings to identify semantically similar literature based on input papers or even unpublished paper abstracts using SPECTER~\cite{cohan2020specter} and GloVe~\cite{pennington2014glove}. It also introduced a corpus of more than 59,000 papers across 38 prominent visualization venues, with summative user feedback demonstrating the utility of the tool.

While prior approaches to visualize academic articles have made significant strides towards finding semantically related articles~\cite{alexander2014serendip,isenberg2016visualization,wang2019vispubcompas,narechania_vitality_2022} and exploring citation networks~\cite{chou2011papervis,heimerl2015citerivers,dattolo2018visualbib,wilkins2015evolutionworks}, these efforts nonetheless are limited in their ability to support intuitive interaction with a corpus of academic literature or assist individuals in summarizing large bodies of literature. We recognize the past two years have seen an explosive growth and innovation in the capability and range of applications of Large Language Models (LLMs)~\cite{zhao2023survey}, including use within the visualization community for generating visualizations~\cite{tian2024chartgpt, sah2024nl4dvllm} and authoring data-driven articles~\cite{sultanum2023datatales}. We observe an opportunity to build upon prior literature search visualization approaches and recent developments with LLMs to introduce literature search methods to address these gaps.

We present \system{}, an open-source visualization system for conducting literature searches, that incorporates a novel Retrieval Augmented Generation (RAG) architecture~\cite{lewis_retrieval-augmented_2020}. \system{} enables users to search for relevant literature using (1) a paper(s) as the seed to find similar work, (2) the abstract of an existing paper or paper to-be-written, (3) traditional keyword-based searching, or (4) a natural language query to an LLM-powered chat interface. Results are queried from a vector database and shown in a rank-ordered table with a similarity score including title, abstract, authors, citation counts, and links to the original sources. Results are also visualized in a projection where papers that appear closer together spatially are more similar in the vector embedding space. Users can also interact with the LLM interface to summarize a corpus of papers and ask contextually relevant questions, e.g., \textit{``what is the grounded theory method?''}

Over the \prevsystem{} predecessor system, \system{} introduces a number of novel features, including its RAG architecture and integration of prompt chaining (described in Section~\ref{sec:architecture}). The system supports natural context-aware conversations through LLMs to summarize and understand collections of papers. The system also augments the dataset of 59,000 papers from \prevsystem{} to cover publications across 38 visualization venues over the past 3 years for a total of 66,692 papers. \system{} is available as open-source software at \url{https://vitality-vis.github.io}.

The remainder of this paper is organized as follows: we review relevant background information in Section~\ref{sec:rw}. Next, we discuss the architecture of \system{} in Section~\ref{sec:architecture} followed by a description of the front-end interface and system capabilities in Section~\ref{sec:system}. We provide exemplary usage scenarios in Section~\ref{sec:scenarios}. 
Finally, we discuss the implications and applications of this work in Section~\ref{sec:discussion}.

\section{Related Work}
\label{sec:rw}

With the ever growing body of scientific literature, there is a constant need for support to conduct more effective and efficient literature research. This challenge has attracted increasing research attention as more publications become digitally available more widely. Early work comes from both within the visualization community (such as PaperVis~\cite{chou2011papervis}) and beyond (such as the work by El-Arini and Guestrin~\cite{el-arini_beyond_2011}). For this paper, we will focus on the results from visualization and related communities. 

Broadly speaking, the visualization research for literature analysis spans a spectrum, from providing an overview of a field to helping find and understand papers related to a specific topic. For the former, a well-known example is the work by Isenberg et al.~\cite{isenberg2016visualization} that aims to provide an overview of the entire visualization research landscape, using papers spanning two and a half decades. While there are other efforts that utilize a similar paper collection, i.e., all the papers from IEEE VIS, few solely focus on the large picture. 

The other end of the spectrum targets literature for a specific topic. This is closer to the goal of \system{}, so we would focus our discussion here. Common approaches employed by these visualization efforts include topic or semantic analysis, network analysis, and supporting the literature review workflow. From early on it has been recognized that keyword matching alone is not an effective way to find relevant papers~\cite{el-arini_beyond_2011, wang2019vispubcompas,literature_explorer}, as a concept or technique can often be expressed in several different ways. Most of the proposed approaches use some form of text analytics or natural language processing methods. Topic modeling is a popular approach among the early work that can help identify relevant papers by grouping similar ones together~\cite{el-arini_beyond_2011, wang2019vispubcompas,literature_explorer}. This includes techniques that are designed for text analysis in general and can be easily applied to literature analysis such as Serendip~\cite{alexander2014serendip}. As NLP research progresses, new techniques are being used for literature analysis, such as text embeddings generated by transformer-based models~\cite{benito-santos_glassviz_2020}, including GloVe~\cite{glove} and SPECTER~\cite{cohan2020specter} used in \prevsystem~\cite{narechania_vitality_2022}. \system{} takes advantage of text embeddings created by the latest LLMs, which represent a breakthrough for many NLP tasks, achieving performance levels close to humans.

Network analysis is another common approach used by literature visualization, creating and visualizing co-author and affiliation network~\cite{wang2019vispubcompas}, citation network~\cite{chou2011papervis,heimerl2015citerivers,yoon_conference_2020}, and similarity network~\cite{chou2011papervis,isenberg2016visualization, benito-santos_glassviz_2020, narechania_vitality_2022}, among others. \system{} focuses on paper similarity, using the latest LLM-based embeddings to create a more accurate network topology. Finally, there are a series of work targeting the literature review workflow, such as LitSense~\cite{sultanum_litsense_2020} and Relatedly~\cite{palani_relatedly_2023}. While not a focus, the interface and interactions of \system{} are designed to match and support the workflow of common literature analysis tasks.

Dimensionality reduction is essential for visualizing high-dimensional data, such as the embeddings derived from LLMs. \system{} leverages state-of-the-art techniques to reduce the dimensionality of embeddings while retaining their semantic relationships. UMAP~\cite{becht2019dimensionality} has been shown to outperform traditional methods like t-SNE~\cite{van2008visualizing} and PCA~\cite{mackiewicz1993principal} in terms of preserving both local and global structures in the data. This capability makes UMAP particularly suitable for visualizing the complex relationships within the literature embeddings generated by LLMs. Once the embeddings are reduced to a lower-dimensional space, they are visualized using interactive plots. These plots allow users to explore the relationships between different papers intuitively.

Using LLMs for literature visualization is still in its infancy. The closest example we found is SciDaSynth~\cite{wang_scidasynth_2024}, which has not been peer reviewed. While it also uses LLMs to create publication embedding and provides a visual interface for exploration and analysis, it does not come with a collection of papers, and the focus of analysis is on paper comparison. Similar to SciDaSynth, \system{} introduces LLMs while retaining the original paper visualization and retrieval capabilities of \prevsystem{}. \system{} offers users a novel natural language-based interaction for paper retrieval. The search results are integrated into the existing exploratory and analytical visualization panels of \prevsystem{}, thereby combining LLMs technology with visualization.

\section{Architecture and Implementation}
\label{sec:architecture}

\begin{figure}[ht]
    \centering
    \includegraphics[width=0.8\columnwidth]{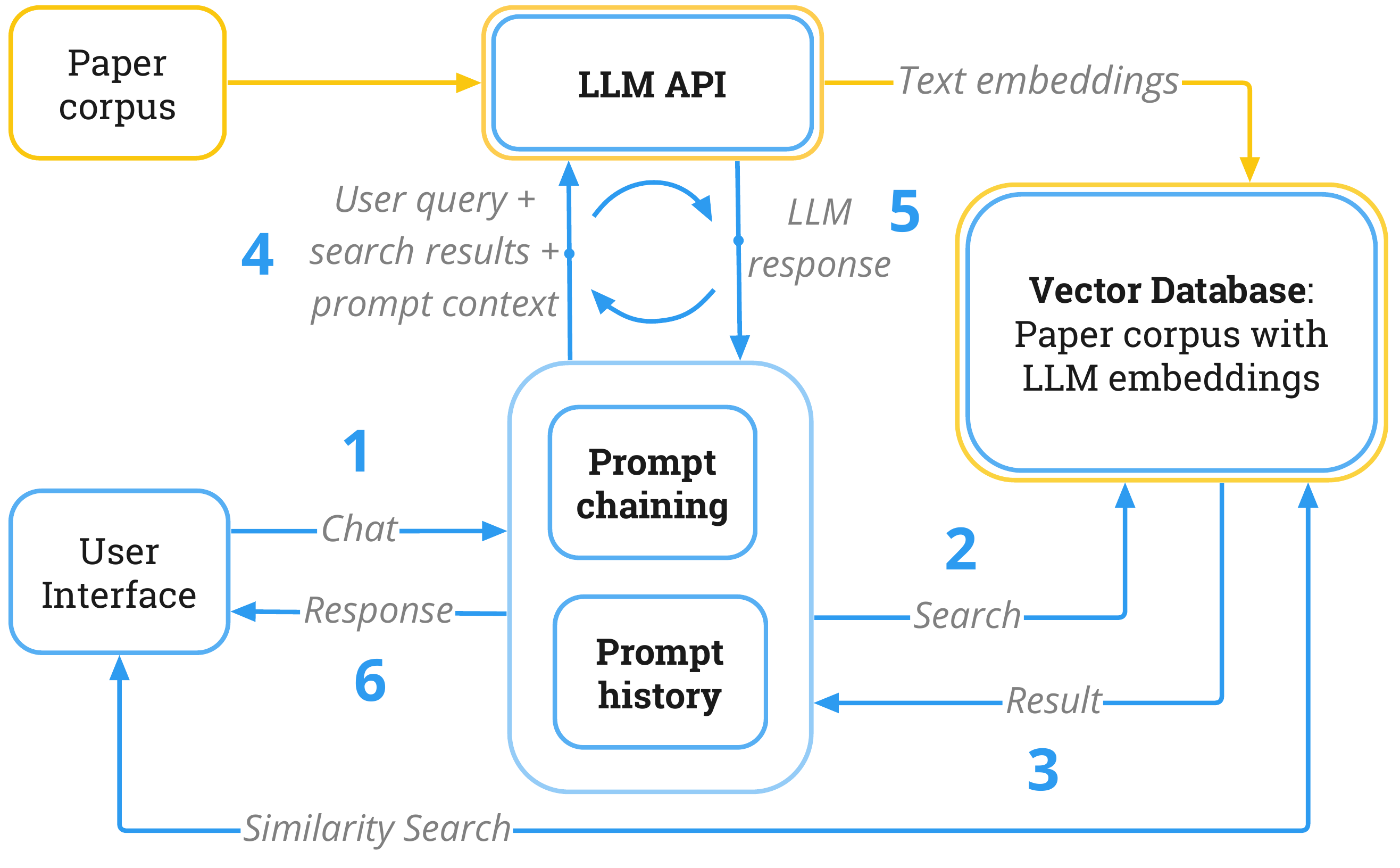}
    \caption{Architecture of \system{}: \textbf{(Step 1)} User input to the system. \textbf{(Step 2 \& 3)} Retrieve data from the vector database. \textbf{(Step 4)} Combine the result with user input in the prompt. \textbf{(Step 5)} Recall result from LLM. \textbf{(Step 6)} Return the final result to the user.}
    \label{fig:arch}
\end{figure}

Figure~\ref{fig:arch} illustrates the architecture, flow of data, and core links of \system{}. 
All the papers in the corpus are first pre-processed using an LLM to create their embeddings (for similarity search), with the results stored in a vector database. 
This is shown as the orange part on the top. Similarity search is then performed within the vector database without further LLM access (the bottom line). User input from the chat interface (Step 1) is broken down into a series of smaller tasks using ``Prompt chaining.'' For each sub-task, relevant information is retrieved from the vector database (Step 2 \& 3) and then combined with user input in the prompt before sending it to the LLM (Step 4). ``Prompt history'' provides context from previous conversations. Once all the sub-tasks are completed (Step 5), the final results are returned to the user (Step 6).

\subsection{Similarity Search using LLM}

All the papers in the corpus are pre-processed using an LLM to create their embeddings based on paper metadata, which includes the title, authors, the conference or journal in which the paper is published, publication date, keywords, and abstract. 
These text embeddings convert textual data into high-dimensional vectors that capture semantic relationships among paper metadata~\cite{tang_pte_2015}, ensuring thorough comprehension of each paper and enabling more effective similarity searches within the vector database.
For clarity, the ``text embedding'' mentioned refers to embeddings created from paper metadata rather than the full paper content. This metadata-focused approach balances detail with computational efficiency and allows for robust semantic matching.

To find similar papers from an academic literature corpus, while \prevsystem{} uses GloVe~\cite{glove} and SPECTER~\cite{cohan2020specter} embeddings,
\system{} uses ADA, the embeddings generated by OpenAI's \textit{text-embedding-ada-020} model~\cite{ada2}, which is a breakthrough model that generates contextually rich embeddings by leveraging advanced attention mechanisms and transformer architectures.  
Thus, \system{} now includes ADA, GloVe, and SPECTER embeddings.
\system{} also leverages two \textit{vector databases} (Faiss~\cite{johnson2019billion} and ChromaDB~\cite{chromadb}) to store and manage these embeddings and also to locate similar papers via approximate nearest neighbor search. 

\subsection{Chat with Papers using RAG and Prompt Chaining}

\begin{figure*}[t]
    \centering
    \includegraphics[width=\linewidth]{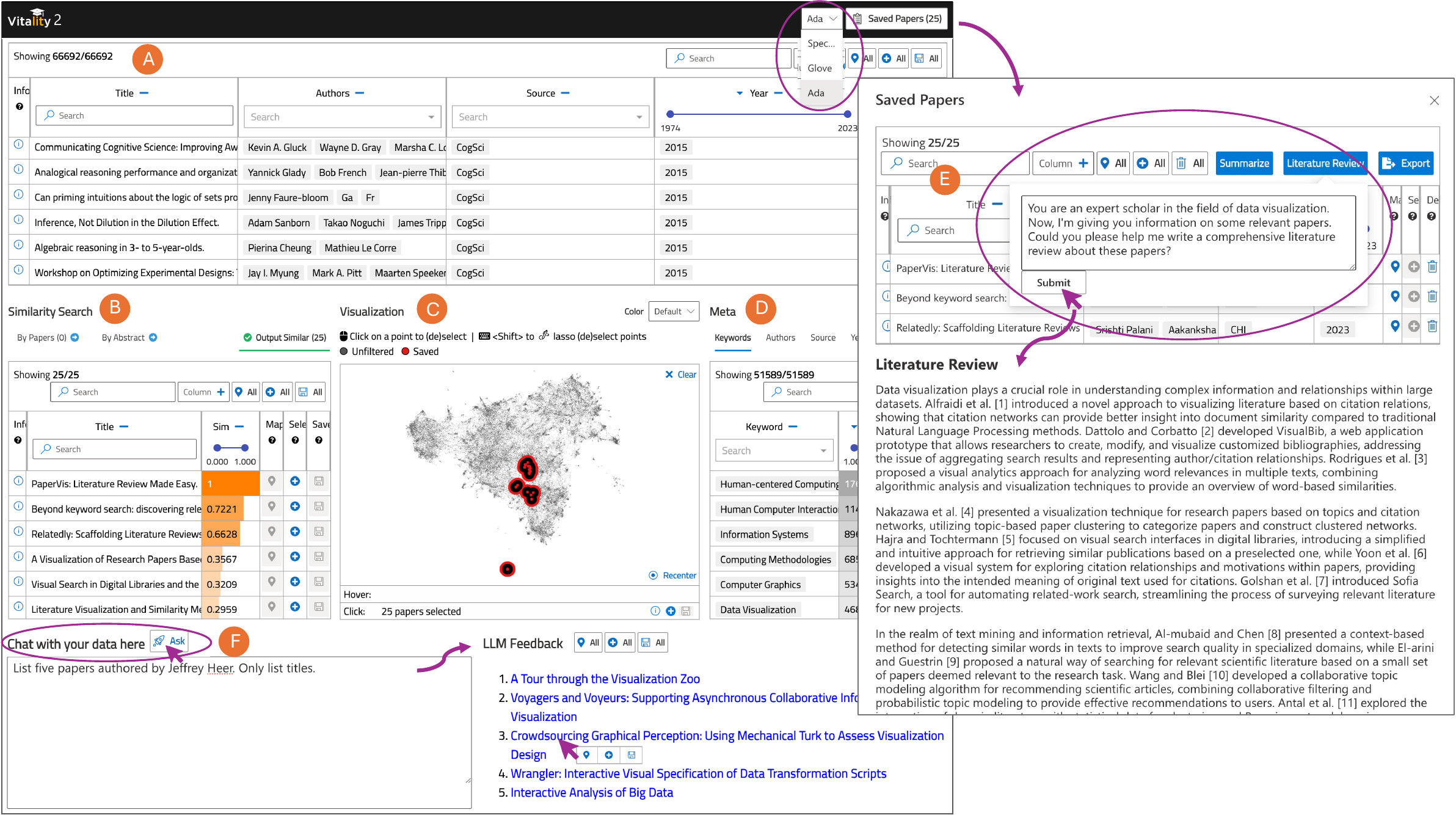}
    \caption{The \system{} User Interface. (A) \textbf{Paper Collection View} shows the entire corpus of publications, (B) \textbf{Similarity Search View} shows options to look-up publications that are similar to another list of publications or by a work-in-progress title and abstract, (C) \textbf{Visualization Canvas} shows an interactive 2-D UMAP projection of the embedding space of the entire paper collection, (D) \textbf{Meta View} shows summaries of certain attributes with respect to the Paper Collection View (A), (E) Opens a \textbf{Saved Papers View} from which the saved papers can be exported as JSON. 
    Extending \prevsystem, we added (F) \textbf{Chat with your Data} view to allow users to ask natural-language based questions based on the paper corpus. 
    We also added ADA embeddings (in addition to GloVe and SPECTER embeddings) and enable users to \emph{Summarize} or write a \emph{Literature Review} on the \textbf{Saved Papers} using LLMs, including the ability to customize the prompts.}
    \label{fig:vitality2}
\end{figure*}

While text embeddings and vector databases allow finding similar papers, there are several other kinds of analyses a user may want to perform, such as understanding a technical concept, summarizing a single paper, or writing a literature review based on multiple papers. To perform such analyses, users may need to build customized tools and/or learn a new query language, neither of which are readily accessible to nor understandable by many users. 
\system{} addresses this issue by leveraging the powerful natural language understanding and generation capabilities of LLMs, allowing users to directly engage with the LLM using natural language. 

However, there are a few challenges when applying LLMs for such analyses: 
(1) \emph{Prompt Size} -- This refers to the largest prompt that an LLM can accept and depends on the LLM type and version. Earlier versions of LLMs accepted a few thousand tokens, wherein each word in a prompt accounted for a few tokens. More recent versions have a larger limit, e.g., up to 16K tokens for GPT 3.5~\cite{gpt-models}. However, this can still limit analyses involving a large corpus of papers, which often have thousands of words each.
(2) \emph{Hallucination} -- it is well known that LLMs can create seemingly plausible information about something that does not exist~\cite{yao_llm_2023}, e.g., suggest papers that never existed, that can be detrimental. 
(3) \emph{Limited Comprehension Capabilities} -- Despite the ability to generate seemingly logical and coherent text, LLMs lack genuine comprehension capabilities. This implies that an LLM may encounter difficulties when handling complex problems or tasks, particularly those requiring a deeper understanding of context or concepts.

\begin{figure*}[t]
    \centering
    \includegraphics[width=\linewidth]{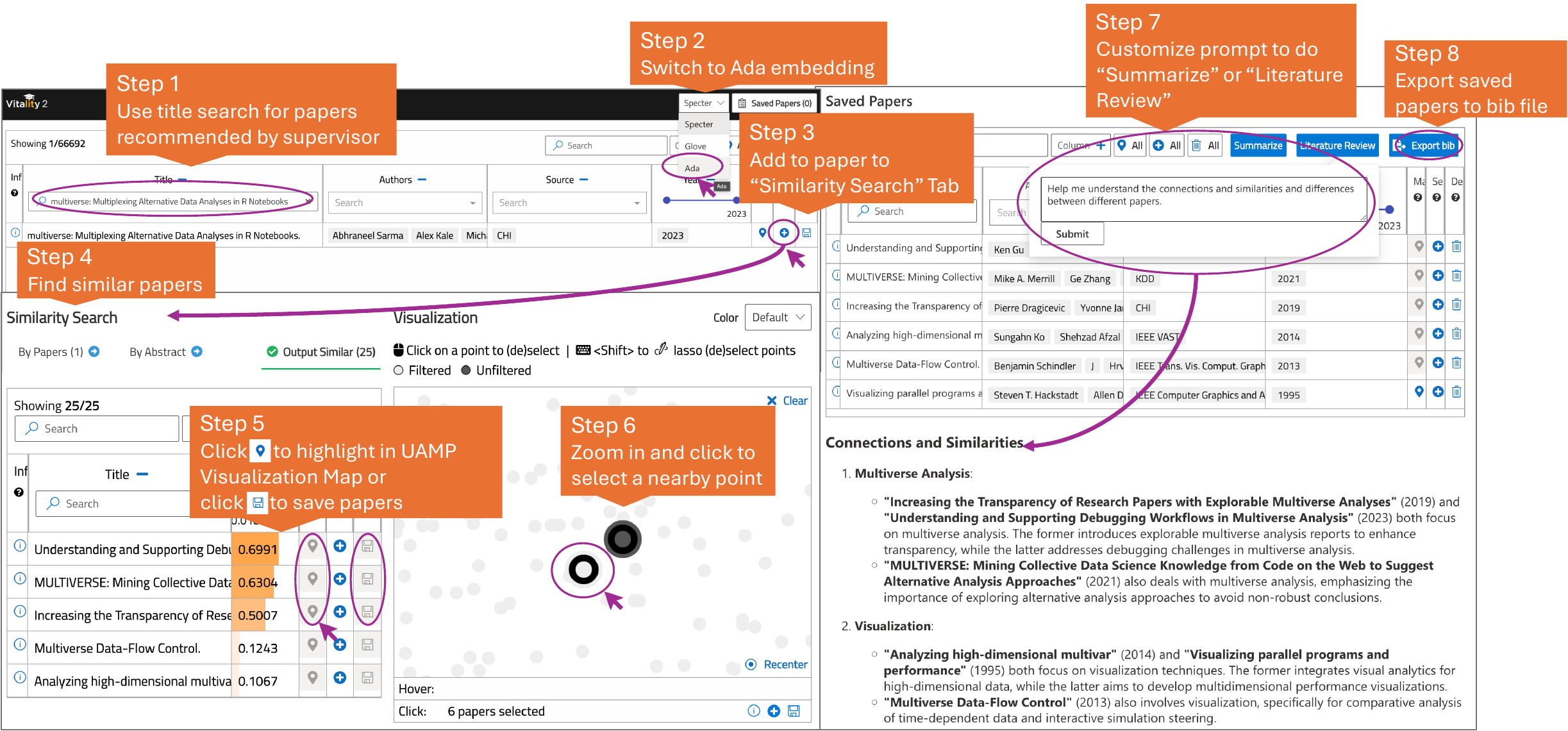}
    \caption{\userN{}'s process of doing literature review using \system{}. \textbf{(Step 1)} \userN{} searches the \system{} database for articles recommended by supervisor. \textbf{(Step 2)} \userN{} selects Ada Embedding as the embedding option used by \system{} in ``Similarity search''. \textbf{(Step 3)} \userN{} adds the paper she just searched as a seed for ``Similarity Search''. \textbf{(Step 4)} \userN{} uses ``Similarity Search'' to find some related papers. \textbf{(Step 5)} \userN{} saves papers with similarity score of $>0.1$ and highlights them in the UMAP Visualization MAP. \textbf{(Step 6)} \userN{} selects an additional paper that interested her in the UMAP visualization map. \textbf{(Step 7)} \userN{} tries to modify and use different prompts and does ``Summarize'' and ``Literature Review''. \textbf{(Step 8)} \userN{} exports the saved papers to a bib file.
    } 
    \label{fig:usage1}
\end{figure*}

\begin{figure*}[t]
    \centering
    \includegraphics[width=\linewidth]{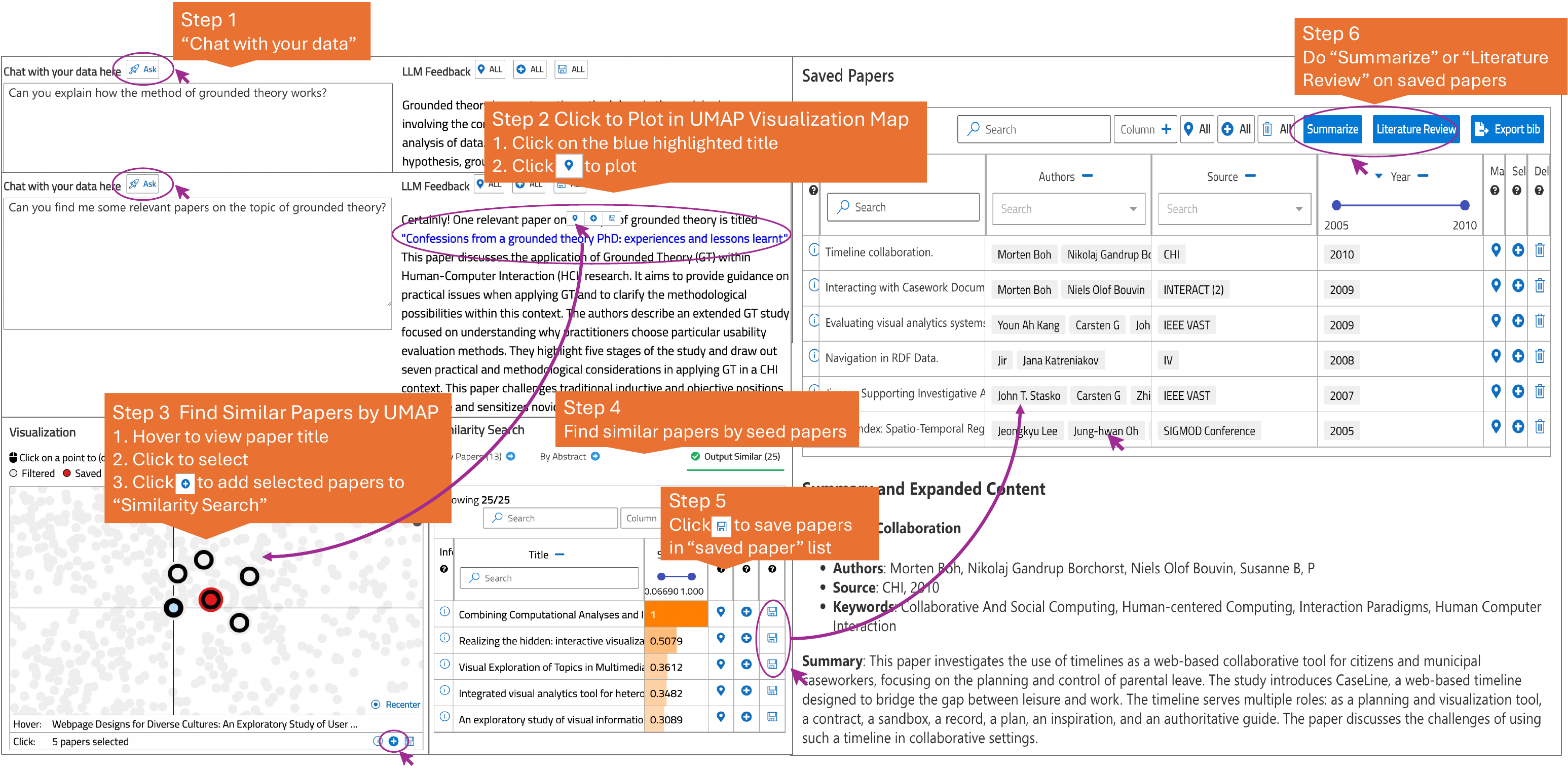}
    \caption{\userA{}'s process for literature search using \system{}. \textbf{(Step 1)} \userA{} utilizes the ``Chat with your data'' feature to quickly explore the domain of ``grounded theory'', an area unfamiliar to him. \textbf{(Step 2)} \userA{} plots an paper cited in the LLM feedback onto the UMAP visualization interface. \textbf{(Step 3)} \userA{} selects a set of closely related papers from the UMAP visualization and \userA{} adds these papers to the ``Similarity Search''.\textbf{(Step 4)} \userA{} uses ``Similarity Search'' feature to find some semantically similar papers. \textbf{(Step 5)} \userA{} saves a subset of papers of particular interest to his ``saved papers'' list. \textbf{(Step 6)} \userA{} employs the ``Summarize'' and ``Literature Review'' feature to review the saved papers.}
    \label{fig:usage2}
\end{figure*}

To overcome these issues, we employ two popular approaches: Retrieval Augmented Generation (RAG) and Prompt Chaining.

\vspace{1mm}\noindent\textbf{RAG}. RAG semantically processes the user's input query to only retrieve relevant information from within the data corpus, thereby reducing the subsequent prompt size and also minimizing the risk of hallucination. 
This retrieval process is not about finding exact answers but rather fetching documents that contain potentially useful context or information related to the query. 

\vspace{1mm}\noindent\textbf{Prompt Chaining}.
Prompt chaining breaks down complex tasks to a series of smaller steps and provides specific prompts that are known to be effective for each of these steps~\cite{wu_ai_2022}.
\system{} borrows this idea and implements a conversation framework to manage the content and context of user queries and LLM responses. \system{} uses LangChain~\cite{langchain} a popular open-source library for this purpose. As LLM API calls operate independently of each other without inherent memorization or management of the context, there is no built-in functionality for retaining conversation history. 
Sending the entire dialogue history with each LLM API call risks exceeding the maximum token limit. 
Therefore, we adopt a two-step approach in \system{}: first, a summary of recent conversations is generated by calling the LLM API once to obtain a ``condensed conversation history.'' 
Then, in the second API call, the user's query, conversation history, and retrieved information are concatenated into a prompt and sent to the LLM API. 
This an example of prompt chaining, that efficiently combines the results of two or more LLM API calls.

\section{\system{}}
\label{sec:system}

We present \system{}, an LLM-powered visual analytics tool to help users write academic literature reviews.

\subsection{Dataset of Academic Articles}

\prevsystem{}~\cite{narechania_vitality_2022} provided a dataset of 59,232 academic papers from visualization and HCI literature, along with their metadata such as their title, abstract, author(s), keyword(s), publisher, publication year, citation counts, and n-dimensional and 2-dimensional vector embeddings (GloVe and SPECTER).
\prevsystem{} also open-sourced a web-scraping framework to extract the above information from digital repositories such as IEEE Xplore and ACM Digital Library for continued development of the corpus. 
\system{} used this scraper to extract more recent papers between 2021-2023, resulting in an augmented dataset of 66,692 papers.

\subsection{User Interface}

Figure~\ref{fig:vitality2} shows the \system{} user interface, illustrating the new features built on top of \prevsystem. 

In \prevsystem{}, the \textbf{Paper Collection View} (A) shows the entire corpus of publications, the \textbf{Similarity Search View} (B) shows options to look-up publications that are similar to another list of publications or by a work-in-progress title and abstract, the \textbf{Visualization Canvas} (C) shows an interactive 2-D UMAP projection of the embedding space of the entire paper collection, the \textbf{Meta View} (D) shows summaries of certain attributes with respect to the Paper Collection View (A), and (E) opens a \textbf{Saved Papers View} from which the saved papers can be exported in a JSON and .bibtex format for later use. We added three new capabilities in \system{}.

First, we expanded available embedding options (from GloVe and SPECTER) to add OpenAI's ADA~\cite{neelakantan2022text}. 
Users can view the 2-dimensional ADA embeddings in the UMAP and search for similar papers (by title or abstract).
Based on our own testing and consistent with prior benchmarks~\cite{muennighoff2022mteb}, we found ADA embeddings to perform better than GloVe and SPECTER on \system{} features.

Second, we added a new \textbf{Chat with your Data View} (F) to allow users to ask natural-language questions based on the entire corpus, e.g., \emph{``Help me find some papers related to geographic science visualization.''}
For papers mentioned in the LLM's output, \system{} allows the user to map (view in the UMAP), select (look up other similar papers), or save them (include it in a literature review or export).
This capability holds immense potential for revolutionizing scholarly information access, enabling users to effectively harness the capabilities of RAG and pose a wide spectrum of inquiries.

Third, we also added novel capabilities to summarize papers and conduct literature reviews (in the \textbf{Saved Papers} view), building upon the robust semantic understanding capabilities of LLM. 
Clicking the ``Summarize'' button outputs a short summary of each of the saved papers, one below the other.
Clicking the ``Literature Review'' button goes a step further, and outputs a comprehensive literature review based on the saved papers, including descriptions, comparisons, and a bibliography.
These enhancements enable users to swiftly grasp the key information contained within their saved papers or draft an early version of their related work section in their ongoing manuscript. 
Note that \system{} allows users to customize the base LLM prompts to control the verbosity, writing style, and format of the LLM's response.

\section{Usage Scenarios}
\label{sec:scenarios}

\subsection{LLM Summarization of Literature Review}

\userN{} is an undergraduate student who has joined a visualization research lab, working under the supervision of a faculty member and Ph.D. student on an ongoing project conducting controlled experiments on multiverse analyses. Per her supervisor's suggestion, she uses \system{} to kickstart her literature review. 

She begins using the ``Saved Papers'' feature in \system{} to collect papers through various search methods. She starts with a recent paper her supervisor suggested from the CHI conference by Sarma and colleagues~\cite{sarma2023multiverse} and both ``Selects'' it and ``Saves'' it. From this selection, she searches for related papers using the ADA embedding. Of the 25 output similar papers, she decides to save any that have a similarity score of $>0.1$, which results in an additional 5 papers (e.g.,~\cite{gu2023understanding,ko2014analyzing}). She highlights the papers in the UMAP visualization and hovers on nearby papers, selecting an additional relevant paper on parallel program performance~\cite{hackstadt1995visualizing}.

\userN{} visits the set of saved papers in ``Saved Papers'' and selects ``Literature Review.'' \system{} shares (1) a paper-by-paper summarization, then (2) a comparison and contrast of the set of 7 total papers.
\userN{} adjusts the default prompt to describe a higher-level summary to suit her experience, then reads through the result. The literature review points out the motivation of most of the papers on transparency of data analysis, and reveals a divide where some papers represented earlier relevant ideas (e.g.,~\cite{ko2014analyzing,schindler2012multiverse}), while the others were more recent (e.g.,~\cite{sarma2023multiverse,merrill2021multiverse}). She accordingly uses these observations to help her write a first draft of the related work section on multiverse visualization to her team's Overleaf document. She uses the ``Export'' feature of \system{} to add the related papers to the .bib file, and she adds a short paragraph detailing her literature review methodology -- including use of \system{} and LLM summarization in the initial draft.

She reads the papers in more detail, iterates on the writing accordingly, then shares it with her supervisor for feedback. She begins her next literature review task: identifying and summarizing literature on general data analysis workflows with \system{}.

\subsection{Contextual Conversations about Papers with LLM}

\userA{} is a new faculty member at a university, having just defended his Ph.D. Having primarily used quantitative methods for his dissertation research, he wants to explore qualitative methods in his next project: conducting in-depth interviews to understand the potential of visualization to address user concerns around misinformation. 

\userA{} begins using \system{} to search for a CSCW paper he knows on misinformation in times of crisis~\cite{huang2015connected}. He selects the ``info'' icon to read the abstract and additional paper metadata. The abstract mentions the use of \textit{``constructivist grounded theory to guide [their] inquiry.''} \userA{} has heard of this method before, but is not familiar with how it works. Using the LLM interface in \system{}, he asks \textit{``can you explain how the method of grounded theory works?''} \system{} responds with an in-depth description (Figure~\ref{fig:usage2}).
He then asks \textit{``can you find me some relevant papers on the topic of grounded theory?''} The system responds with a description of some relevant papers, which use the grounded theory method. The result is shown in Figure~\ref{fig:usage2}.

\userA{} continues his literature review accordingly, having a better baseline understanding of the method. \userA{} notices that \system{} highlights the titles of the papers cited in the answer in bold blue text. \userA{} clicks on these titles, and \system{} pops up a function box, including (1) Highlight in the UMAP visualization, (2) Add the paper to the similarity search, and (3) Save the paper to the ``Saved Papers'' list. \userA{} selects one of the papers he is interested in and uses the similarity search function to find other papers similar to it. Next, \userA{} saves all of these articles in the ``Saved Paper'' list. 
Finally, \userA{} uses the \textit{Summarize} and \textit{Literature Review} features to further review on these saved papers.

\section{Discussion, Limitations, \& Future Work}
\label{sec:discussion}

We observe a few limitations to our current approach. First, while using LLMs to summarize a set of papers can be a useful starting point for a literature review, the quality of the output is far from sufficient for being included in a paper as-is. The summarization is based on the meta-data that is contained in the \system{} paper corpus, which does not include full-text of the articles. A potential solution is to optimize the web crawler to also retrieve the full text of papers, segment these texts into appropriately sized chunks~\cite{yepes2024financial} and generate embeddings for these chunks. These embeddings, along with the associated meta-data (title, authors, keywords, abstract, etc.), should then be stored in the vector database. This will ensure more comprehensive and accurate summaries.

Furthermore, some studies also indicate that there is concern within the academic community regarding the development and use of LLMs~\cite{KIM2023598}. The accuracy of response by LLMs depends on the initial training. Potential biases, inaccuracies, or misunderstandings present in the data used for training the models can lead to erroneous outputs. To mitigate this problem, we suggest adding a user prompt in \system{} to explicitly warn users about the limitations of content generated by \system{}, thereby reminding and cautioning users to use this tool with care. 

Another concern is hallucination from LLMs~\cite{xu2024hallucination}. In one instance, when asked to describe the concept of ``grounded theory'' in Usage Scenario 2 (Section~\ref{sec:scenarios}), the LLM described a reasonable high-level summary of the approach, but referenced a paper not contained in the \system{} database. Upon searching the title, it referred to a reasonably popular book from Birks and Mills in 2015 with more than 2,000 citations at the time of this writing~\cite{birks2015grounded}. However, when asking for additional relevant papers on the topic of grounded theory (implying that they come from the \system{} corpus), the LLM responded with some paper titles that were neither contained in the \system{} database, nor obvious paper titles outside the database according to a Google Scholar search. 
Although completely eliminating hallucinations in LLMs remains challenging~\cite{yao2023llm}, various strategies exist to reduce their occurrence. As discussed in this paper, actually the RAG architecture is one such approach. RAG mitigates LLMs hallucinations by providing additional contextual information. In future work, we can further reduce hallucinations in RAG by incorporating external knowledge bases from reliable sources, such as Google Scholar search. Additionally, as research on LLMs advances, several potential methods to address the issue of hallucinations in LLMs have emerged. 
Recent benchmarks suggest that the incidence of AI hallucination is relatively small for GPT-4 by OpenAI~\cite{hallucination_test_2023}.

\section{Conclusion}
We introduced a system, \system{}, for conducting literature review using a RAG architecture, a corpus of more than 66,000 papers from visualization-related venues, and novel features for interacting with the paper corpus through Large Language Models (LLMs). We provide the system as an open-sourced code contribution at \url{https://vitality-vis.github.io}, alongside the paper corpus, and hope to stimulate future work in optimizing literature review methods.

\bibliographystyle{abbrv-doi}

\bibliography{references}

\end{document}